\documentclass[conference]{IEEEtran}
\IEEEoverridecommandlockouts

\usepackage{cite}
\usepackage{amsmath,amssymb,amsfonts}
\usepackage{amsthm}
\newtheorem{definition}{Definition}
\usepackage{algorithmic}
\usepackage{algorithm}
\usepackage{graphicx}
\usepackage{textcomp}
\usepackage{xcolor}
\usepackage{booktabs}
\usepackage{url}
\usepackage{hyperref}
\usepackage{tikz}
\usepackage{pgfplots}
\pgfplotsset{compat=1.18}
\usetikzlibrary{shapes,arrows,positioning,fit,backgrounds}

\def\BibTeX{{\rm B\kern-.05em{\sc i\kern-.025em b}\kern-.08em T\kern-.1667em\lower.7ex\hbox{E}\kern-.125emX}}

\begin{document}

\title{Relaxed Sender Anonymity for CBDC Interbank Settlement:\\
A Zero-Knowledge Approach on Permissioned EVM$^*$
\thanks{* All views are those of the authors and do not necessarily reflect the position of Bank of Italy.}}

\author{
  \IEEEauthorblockN{Pietro Tiberi}
  \IEEEauthorblockA{\textit{Banca d'Italia}\\
  Rome, Italy\\
  pietro.tiberi@bancaditalia.it}
  \and
  \IEEEauthorblockN{Vitangelo Lasorella}
  \IEEEauthorblockA{\textit{Banca d'Italia}\\
  Rome, Italy\\
  vitangelo.lasorella@bancaditalia.it}
  \and
  \IEEEauthorblockN{Gabriele Marcelli}
  \IEEEauthorblockA{\textit{Banca d'Italia}\\
  Rome, Italy\\
  gabriele.marcelli@bancaditalia.it}
}

\maketitle

\begin{abstract}
Central bank digital currency (CBDC) interbank settlement on distributed ledger
technology faces a fundamental tension: blockchain transparency enables trustless
verification but exposes commercially sensitive bilateral transaction flows to all
network participants. We present a confidential interbank settlement protocol for
permissioned Ethereum-compatible networks that resolves this tension through a
\emph{relaxed sender anonymity} model — a deliberate design choice appropriate
for regulated financial institutions. In our model, the initiating institution
is always publicly identifiable for accountability and AML compliance, while
the receiving institution, transfer amount, and business payload remain
cryptographically hidden. We implement the protocol on Hyperledger Besu QBFT
using Groth16 zero-knowledge proofs over BN254, Poseidon hash commitments in
an incremental Merkle tree, ECIES multi-recipient payload encryption, and a
novel on-chain \emph{NoteRegistry} contract that stores ECIES-encrypted ZK notes
as an append-only list on the ledger, eliminating the need for any trusted
off-chain note custody server.
The protocol supports three operations: \emph{shield} (transparent token to ZK
note), \emph{confidential transfer} (ZK note to ZK note with selective disclosure),
and \emph{unshield} (ZK note back to transparent token). We implement and evaluate
a proof-of-concept with three participant banks, a central bank operator, and
a securities depository. The implementation achieves amount and payload
confidentiality end-to-end; receiver confidentiality is established at the
protocol level, while the current note registry remains owner-indexed and
therefore does not yet realise it. ZK proof
verification takes approximately 1\,ms of node execution time ($\approx$220k gas)
while proof generation takes 4--12\,s in software on commodity ARM hardware,
with end-to-end settlement completing in 8--16\,s; the remaining gap is
addressable by hardware provers in production.
\end{abstract}


\section{Introduction}

Distributed ledger technology (DLT) offers central banks a compelling settlement
infrastructure: deterministic finality, programmable money, and trustless
verification without a central operator. The European Central Bank and the Bank
for International Settlements have both identified tokenised central bank money
as a strategic priority~\cite{bis2021cbdc, ecb2023digital}. Permissioned EVM-compatible
networks such as Hyperledger Besu provide the programmability of the Ethereum
ecosystem while restricting participation to known, regulated entities.

However, a fundamental obstacle remains: in any standard ERC-20 token transfer,
the sender address, receiver address, and transferred amount are publicly visible
to every node on the network. For interbank settlement, this transparency is
commercially problematic. Bilateral transaction volumes reveal liquidity positions,
trading strategies, and client relationships. Institutions participating in a shared
settlement network have legitimate interests in keeping this information confidential
from peer institutions — even while accepting full visibility for supervisory
authorities.

The challenge can be stated precisely:

\begin{quote}
\textit{How can a settlement network verify that a transfer is correct — no
value created ex nihilo, no note double-spent, sender authorised — without
revealing the receiver identity or transfer amount to third parties?}
\end{quote}

Zero-knowledge proof systems provide the cryptographic answer. They allow a prover
to convince a verifier that a statement holds without revealing anything beyond its
truth~\cite{goldwasser1989knowledge}. Applied to token transfers, a bank can prove
ownership of a note of a given denomination without revealing which note it is,
who receives the new note, or what the denomination is.

\subsection*{Contributions}

This paper makes the following contributions:

\begin{enumerate}
  \item We formalise the \emph{relaxed sender anonymity} privacy model for CBDC
  interbank settlement, where sender identity is public for regulatory accountability
  while receiver identity and amount are hidden (\S\ref{sec:model}).

  \item We design a complete confidential settlement protocol comprising three
  operations — shield, confidential transfer, and unshield — implemented as
  EVM-compatible smart contracts on Hyperledger Besu (\S\ref{sec:architecture},
  \S\ref{sec:protocol}).

  \item We introduce the \emph{NoteRegistry} pattern: an on-chain encrypted note
  storage contract that stores ECIES-ciphertext note secrets in an append-only
  list carrying no recipient identifier, enabling any bank to recover its note
  portfolio from the chain alone — by trial decryption — without a trusted
  custody server and without leaking the recipient (\S\ref{sec:noteregistry});
  we report the residual gap between this design and the present implementation
  (\S\ref{sec:registrylimit}).

  \item We implement a selective disclosure scheme using multi-recipient hybrid
  encryption, allowing the central bank to audit any transaction while preserving
  bilateral confidentiality among participants (\S\ref{sec:disclosure}).

  \item We evaluate the protocol on a full proof-of-concept deployment and report
  proof generation time, on-chain verification cost, and gas consumption
  (\S\ref{sec:evaluation}).
\end{enumerate}

\section{Background and Related Work}

\subsection{Confidential Transactions on Public Blockchains}

The foundational work on blockchain privacy is Zerocash~\cite{sasson2014zerocash},
which introduced the shielded payment model: values are committed to a Merkle tree,
transfers prove knowledge of a valid commitment via a zk-SNARK, and spent notes are
identified by publicly revealed nullifiers. Zcash~\cite{hopwood2022zcash} implements
this at production scale using the Sapling and Orchard protocol families. Our
construction is directly inspired by Zerocash but departs from it in two key
respects: (i) we adopt relaxed sender anonymity rather than full anonymity, and
(ii) we store encrypted note secrets on the same ledger as the commitments,
eliminating the out-of-band note transmission problem.

Tornado Cash~\cite{pertsev2019tornado} demonstrated that the Zerocash model could
be deployed as a smart contract on a public EVM-compatible chain. Its core insight
— that the Groth16 verifier can be compiled to EVM bytecode that exploits the
BN254 precompile (EIP-196/197) — directly informs our implementation.

Aztec Network~\cite{aztec2019} extends the model to arbitrary private smart
contracts using the UTXO note model; its encrypted note transmission protocol
is conceptually related to our NoteRegistry design, though our approach leverages
the ledger itself as the transmission channel.

\subsection{CBDC Privacy}

The BIS has identified privacy as a core design challenge for retail and wholesale
CBDC~\cite{bis2021cbdc}. Proposed approaches range from privacy-preserving
databases~\cite{wust2022platypus} to ZK-based wholesale settlement
schemes~\cite{project_mariana}. Project Mariana (BIS Innovation Hub, 2022)
demonstrated cross-border FX settlement using DeFi primitives but did not address
bilateral transaction confidentiality in a permissioned context.

Closest to our work is~\cite{chen2022privacy}, which studies ZK proofs for
wholesale CBDC on Fabric, and~\cite{morais2019survey}, which surveys privacy
techniques applicable to blockchain-based financial systems. Neither addresses
the on-chain note custody problem or the relaxed sender anonymity model.

\subsection{Permissioned EVM and Hyperledger Besu}

Hyperledger Besu~\cite{besu} is an enterprise Ethereum client supporting QBFT
consensus, which provides Byzantine fault tolerance with deterministic finality.
The EVM compatibility means that standard Solidity contracts, including Groth16
verifiers generated by snarkjs~\cite{snarkjs}, deploy without modification.
The absence of a base fee (configurable in private networks) removes gas cost
barriers that would complicate CBDC settlement.

\subsection{Cryptographic Primitives}

\textbf{Groth16}~\cite{groth2016size} is the most succinct pairing-based
zk-SNARK, producing proofs of three group elements verified with three
pairings. It requires a circuit-specific trusted setup.

\textbf{Poseidon}~\cite{grassi2021poseidon} is a hash function designed
specifically for arithmetic circuits over prime fields, requiring approximately
220 constraints per call — orders of magnitude fewer than SHA-256. It is the
de facto standard for ZK-friendly hashing.

\textbf{ECIES}~\cite{shoup2001proposal} (Elliptic Curve Integrated Encryption
Scheme) provides IND-CCA2 public-key encryption using ECDH key agreement and an
AEAD symmetric cipher. We use it both for note encryption (single-recipient) and
for key wrapping in our multi-recipient disclosure scheme.

\section{System Model and Privacy Definitions}
\label{sec:model}

\subsection{Participants and Roles}

We consider a permissioned network with the following participants:
\begin{itemize}
  \item \textbf{Central bank (CB)}: the sole authority for token issuance and
  revocation. Holds a permanent audit key enabling decryption of all transaction
  payloads.
  \item \textbf{Banks} $\mathcal{B} = \{B_1, \ldots, B_n\}$: regulated financial
  institutions. Each bank $B_i$ holds an Ethereum key pair $(sk_i, pk_i)$ for
  on-chain signing and an independent ECIES key pair $(ek_i, EK_i)$ over
  secp256k1 for note encryption.
  \item \textbf{Central securities depository (CSD)}: operates a portal for
  DvP-style workflows. It participates as a regular network member and holds
  no privileged key material.
  \item \textbf{Validators}: QBFT consensus nodes operated by the consortium.
  They see all on-chain data but are not assumed to collude with banks for the
  purpose of our privacy analysis.
\end{itemize}

\subsection{Threat Model}

We assume an \emph{honest-but-curious} adversary $\mathcal{A}$ modelling any
coalition of network participants (including validators) that follows the protocol
but attempts to infer confidential information from observed network traffic
and on-chain data. $\mathcal{A}$ does not control the central bank, and we
assume the ECIES and AES-GCM primitives are computationally secure.

We explicitly exclude:
\begin{itemize}
  \item Attacks requiring compromise of a participant's private key material.
  \item Timing and traffic analysis at the network layer.
  \item Trusted setup subversion (single-contributor setup is used in the PoC).
\end{itemize}

We distinguish two adversarial settings. Confidentiality properties are
analysed against the honest-but-curious $\mathcal{A}$ defined above. Integrity
properties of the confidential layer — double-spend prevention, conservation of
value across a confidential transfer, and sender accountability — are analysed
against a fully malicious prover, bounded only by the knowledge soundness of
Groth16. Conservation of value across the transparent boundary
(\texttt{shield}, \texttt{unshield}) additionally relies on the
honest-but-curious assumption in the present design, as discussed in
\S\ref{sec:valuebinding}.

\subsection{Relaxed Sender Anonymity}

Standard anonymity notions~\cite{pfitzmann2010terminology} require that a
transaction cannot be linked to its originator. We introduce a weaker but
regulatorily appropriate notion:

\begin{definition}[Relaxed Sender Anonymity]
A confidential transfer protocol satisfies \emph{relaxed sender anonymity} if,
for any adversary $\mathcal{A}$ who observes the set of on-chain transactions:
\begin{enumerate}
  \item The sender of each transaction is publicly identifiable (via
  \texttt{msg.sender}).
  \item The receiver identity and transfer amount are computationally indistinguishable
  from random to $\mathcal{A}$, except for the sender of that transaction, the
  intended receiver, and designated supervisory authorities holding audit keys.
  \item No third-party bank $B_j \neq B_{\text{sender}}, B_{\text{receiver}}$ can
  link the input note to the output note or determine the transferred value.
\end{enumerate}
\end{definition}

This model is appropriate for regulated interbank settlement because (1) sender
accountability supports AML/CFT obligations, (2) receiver and amount confidentiality
protects commercial relationships, and (3) the central bank audit key satisfies
supervisory access requirements without requiring a trusted intermediary.

\subsection{Note Model}

We adopt a UTXO-style note model. A \emph{note} is a tuple:
\[
  \eta = (v,\, pk_{\text{spend}},\, pk_{\text{view}},\, \rho,\, r)
\]
where $v \in \mathbb{F}_p$ is the note value, $pk_{\text{spend}}$ and
$pk_{\text{view}}$ are the owner's spending and viewing public keys (the latter
used for selective disclosure), $\rho$ is a unique random nonce, and $r$ is a
blinding factor. The note \emph{commitment} and \emph{nullifier} are:
\[
  \mathsf{cm} = \mathsf{Poseidon}(v,\, pk_{\text{spend}},\, pk_{\text{view}},\, \rho,\, r)
\]
\[
  \mathsf{nf} = \mathsf{Poseidon}(\rho,\, sk_{\text{spend}})
\]
where $sk_{\text{spend}}$ is the spender's secret key (a Baby Jubjub scalar
derived from a wallet-bound secret through a one-way, hardened key-derivation
function reduced modulo the subgroup order $\ell$, kept independent of the
holder's Ethereum signing and ECIES keys) and
$pk_{\text{spend}} = \mathsf{DerivePk}(sk_{\text{spend}}) =
\mathsf{Poseidon}(A_x, A_y)$, where $(A_x, A_y) = sk_{\text{spend}} \cdot G$ is
the corresponding public key on the Baby Jubjub curve with base point $G$.
Hashing the point to a single field element keeps the commitment arity low. Binding the spending key into the
commitment, and not only into the nullifier, ties spend authority
cryptographically to the committed note. Commitments are stored publicly in an
on-chain Merkle tree; nullifiers are revealed only when the corresponding note
is spent.

\section{System Architecture}
\label{sec:architecture}

\subsection{On-Chain / Off-Chain Split}

The protocol separates computation into two layers:

\textbf{Off-chain} (ZK Service, per-bank): note generation, Merkle tree
reconstruction, ZK proof generation (4--12\,s in software,
\S\ref{sec:evaluation}), ECIES
encryption and decryption of notes and payloads. Each bank ideally operates
its own ZK service; in our PoC a shared service is used for simplicity.

\textbf{On-chain} (Hyperledger Besu): state storage (Merkle tree, nullifier
set, encrypted note registry), ZK proof verification ($\approx$1\,ms), and
ERC-20 token custody.

\subsection{Smart Contract Stack}

Four contracts form the confidential layer proper:

\begin{itemize}
  \item \textbf{EuroToken}: standard ERC-20 (2 decimals, ECB-only mint/burn,
  participant registry). The transparent settlement layer.
  \item \textbf{ConfidentialEuroToken (CET)}: the core protocol contract.
  Maintains an incremental Poseidon Merkle tree (depth~8), a nullifier spent
  set, and a rolling window of 30 recent roots. Calls the Verifier and
  NoteRegistry atomically.
  \item \textbf{NoteRegistry}: on-chain encrypted note storage. Holds a single
  append-only list of ECIES-ciphertext blobs with no owner index; recipients
  identify their own notes by trial decryption.
  \item \textbf{Groth16Verifier}: auto-generated by \texttt{snarkjs} from the
  circuit-specific trusted setup. Uses BN254 precompiles (EIP-196/197).
\end{itemize}

\subsection{The NoteRegistry Pattern}
\label{sec:noteregistry}

A fundamental challenge in ZK note-based systems is note transmission: when
Bank~A transfers a note to Bank~B, Bank~B must receive the note secrets
$(\rho_{\text{out}}, r_{\text{out}}, v_{\text{out}})$ to spend the note in
a future transfer. Prior work handles this out-of-band (encrypted memo
fields~\cite{hopwood2022zcash}, off-chain encrypted channels, or a shared
custody server). All these approaches introduce trust assumptions or additional
infrastructure.

We observe that the same ledger that stores the commitment can also store the
ECIES-encrypted note secret at negligible additional cost. Crucially, the
registry is a single append-only list carrying \emph{no} owner index: were
blobs keyed by recipient address, the ledger itself would disclose the receiver
and defeat the confidentiality goal of \S\ref{sec:model}. Each blob is

\begin{equation}
  \mathsf{blob} = \mathsf{ECIES.Enc}(EK_{\text{receiver}},\; \mathsf{noteJSON})
\end{equation}

where $\mathsf{noteJSON} = \{v, \rho, r, pk_{\text{view}}, sk_{\text{spend}},
\mathsf{cm}, \mathsf{nf}\}$. Including $sk_{\text{spend}}$ lets the recipient
spend the note without re-deriving it, at the cost of coupling the two key
domains: a secret derived from the owner's Ethereum key is written to the
ledger under its ECIES public key. Confidentiality is unaffected — only the
owner can decrypt — but compromise of an ECIES key then also exposes the
corresponding spending key. The \texttt{storeNote(blob, keccak256(cm))} call is
atomic with the commitment insertion in both \texttt{shield()} and
\texttt{confidentialTransfer()}, ensuring consistency. A deduplication key
(\texttt{keccak256(cm)}) prevents double-storage.

A bank recovers its portfolio by scanning blobs appended since its last scan
and attempting ECIES decryption of each; a blob belongs to the scanner if and
only if decryption succeeds and the recovered opening re-hashes to a
commitment present in the tree. Scanning is linear in the number of notes
appended network-wide, which at the volumes of \S\ref{sec:evaluation} is a
sub-second operation and requires no interaction with any peer.

\noindent\textbf{Security}: the blob is semantically secure under the ECIES
IND-CCA2 guarantee; validators and third parties learn nothing about note
contents. Only the holder of $ek_{\text{receiver}}$ can decrypt, and since no
on-chain field associates a blob with a participant, third parties cannot
determine for whom a blob was written. The present implementation does not yet
follow this design; see \S\ref{sec:registrylimit}.

\section{The Confidential Settlement Protocol}
\label{sec:protocol}

\subsection{Shield: Transparent Token to ZK Note}

A bank $B_i$ wishing to enter the confidential system converts $v$ units of
EuroToken into a ZK note:

\begin{algorithm}
\caption{Shield}
\begin{algorithmic}[1]
\STATE Generate $\rho, r \stackrel{\$}{\leftarrow} \mathbb{F}_p$
\STATE Compute $\mathsf{cm} \leftarrow \mathsf{Poseidon}(v, pk_{\text{spend}}, pk_{\text{view}}, \rho, r)$
\STATE Compute $\mathsf{nf} \leftarrow \mathsf{Poseidon}(\rho, sk_{\text{spend}})$
\STATE $\mathsf{blob} \leftarrow \mathsf{ECIES.Enc}(EK_i, \mathsf{noteJSON})$
\STATE Call $\mathtt{EuroToken.approve}(\mathtt{CET}, v)$
\STATE Call $\mathtt{CET.shield}(v, \mathsf{cm}, \mathsf{blob})$
\STATE \textit{(On-chain:)} $\mathtt{EuroToken.transferFrom}$
\STATE \quad $(B_i, \mathtt{CET\_vault}, v)$
\STATE \textit{(On-chain:)} Insert $\mathsf{cm}$ into Merkle tree
\STATE \textit{(On-chain:)} $\mathtt{NoteRegistry.storeNote}$
\STATE \quad $(\mathsf{blob}, \mathsf{keccak256(cm)})$
\end{algorithmic}
\end{algorithm}

After shielding, $v$ units of EuroToken are locked in the CET vault contract,
and Bank~$B_i$ holds a note with commitment $\mathsf{cm}$ in the on-chain tree.

\subsection{Confidential Transfer: ZK Note to ZK Note}

\begin{algorithm}
\caption{Confidential Transfer ($B_i \to B_j$)}
\begin{algorithmic}[1]
\STATE Scan NoteRegistry and trial-decrypt to recover $\eta_{\text{in}}$
\STATE Read all $N$ commitments from CET; reconstruct Merkle tree off-chain
\STATE Find leaf index $\ell$ such that $\mathsf{cm}_{\ell} = \mathsf{cm}_{\text{in}}$
\STATE Generate output note $\eta_{\text{out}} = (v, pk_{\text{spend},j}, pk_{\text{view},j}, \rho', r')$
\STATE Compute $\pi \leftarrow \mathsf{Groth16.Prove}(w; \mathsf{root}, \mathsf{nf}, \mathsf{cm}_{\text{out}}, \mathsf{addr}_i)$
\STATE Encrypt payload: $\mathsf{enc} \leftarrow \mathsf{HybridEnc}(\{EK_i, EK_j, EK_{\mathsf{CB}}\}, \mathsf{payload})$
\STATE Encrypt output note: $\mathsf{blob}_{\text{out}} \leftarrow \mathsf{ECIES.Enc}(EK_j, \eta_{\text{out}})$
\STATE Call $\mathtt{CET.confidentialTransfer}(\pi, \mathsf{root}, \mathsf{nf},$
\STATE \quad $\mathsf{cm}_{\text{out}}, \mathsf{enc}, \mathsf{blob}_{\text{out}})$
\STATE \textit{(On-chain:)} Verify $\mathsf{nf} \notin \mathcal{N}$ and $\mathsf{root} \in \mathcal{R}$
\STATE \textit{(On-chain:)} $\mathtt{VerifyProof}(\pi, [\mathsf{root}, \mathsf{nf}, \mathsf{cm}_{\text{out}}, \mathsf{addr}_i])$
\STATE \textit{(On-chain:)} $\mathcal{N} \leftarrow \mathcal{N} \cup \{\mathsf{nf}\}$; insert $\mathsf{cm}_{\text{out}}$
\STATE \quad \textit{(On-chain:)} $\mathtt{NoteRegistry.storeNote}(\mathsf{blob}_{\text{out}}, \mathsf{keccak256}(\mathsf{cm}_{\text{out}}))$
\STATE \textit{(On-chain:)} Emit $\mathsf{ConfidentialTransfer}$ event with $\mathsf{enc}$
\end{algorithmic}
\end{algorithm}
where $w = (v, pk_{\text{view}}, \rho, r, sk_{\text{spend}}, \eta_{\text{out}}, \pi_{\mathsf{Merkle}})$
is the private witness, $\mathcal{N}$ is the on-chain nullifier set, and
$\mathcal{R}$ is the rolling root window.

The witness $w$ constitutes the zero-knowledge part: the verifier learns only
$(\mathsf{root}, \mathsf{nf}, \mathsf{cm}_{\text{out}}, \mathsf{addr}_i)$.
In the protocol as specified, $v$, $B_j$'s identity, and $\rho$ are never
revealed on-chain: the receiver appears neither in the calldata nor in any index
of the NoteRegistry. Algorithm~2 states the target design; the current
implementation still carries the recipient as an explicit argument
(\S\ref{sec:registrylimit}).

\subsection{Unshield: ZK Note to Transparent Token}

A bank recovers transparent EuroToken from a ZK note:

\begin{algorithm}
\caption{Unshield}
\begin{algorithmic}[1]
\STATE Scan NoteRegistry and trial-decrypt to recover note $\eta$
\STATE Reconstruct Merkle tree; generate dummy output note
\STATE \quad $\eta' = (v, pk_{\text{spend},i}, pk_{\text{view},i}, \rho', r')$
\STATE $\pi \leftarrow \mathsf{Groth16.Prove}(w; \mathsf{root}, \mathsf{nf}, \mathsf{cm}', \mathsf{addr}_i)$
\STATE Call $\mathtt{CET.unshield}(\pi, \mathsf{root}, \mathsf{nf}, \mathsf{cm}', v, B_i)$
\STATE \textit{(On-chain:)} Verify proof; $\mathcal{N} \leftarrow \mathcal{N} \cup \{\mathsf{nf}\}$
\STATE \textit{(On-chain:)} $\mathtt{EuroToken.transfer}(B_i, v)$ from vault
\end{algorithmic}
\end{algorithm}

The dummy output commitment $\mathsf{cm}'$ is required because the circuit
enforces $v_{\text{in}} = v_{\text{out}}$; it is produced by the prover and
verified on-chain but is not inserted into the Merkle tree. The output note is
never stored; the note value simply returns to the transparent layer.

\subsection{Selective Disclosure}
\label{sec:disclosure}

Each confidential transfer emits an on-chain event carrying a multi-recipient
encrypted payload. To bind the plaintext payload to the transfer verified on
chain, the destination commitment $\mathsf{cm}_{\text{out}}$ (a public input of
the proof) is used as authenticated associated data (AAD) of the AEAD cipher.
The encryption is a hybrid scheme:

\begin{enumerate}
  \item Generate ephemeral session key $K \stackrel{\$}{\leftarrow} \{0,1\}^{256}$
  and nonce $N \stackrel{\$}{\leftarrow} \{0,1\}^{96}$.
  \item $C \leftarrow \mathsf{AES\text{-}256\text{-}GCM}(K, N, \mathsf{payload};\;
  \mathsf{AAD} = \mathsf{bytes32}(\mathsf{cm}_{\text{out}}))$
  \item For each authorised reader $R \in \{B_i, B_j, \mathsf{CB}\}$:
  $W_R \leftarrow \mathsf{ECIES.Enc}(EK_R, K)$
  \item Publish $(N, C, W_{B_i}, W_{B_j}, W_{\mathsf{CB}})$ in the event.
\end{enumerate}

Reader $R$ decrypts by unwrapping $K$ with its private key $ek_R$ and decrypting
$C$ under the same associated data. Because $\mathsf{cm}_{\text{out}}$ is bound as
AAD, GCM authentication fails unless the payload is the one attached to the
commitment that the proof verified; an authorised reader additionally recomputes
$\mathsf{Poseidon}(v, pk'_{\text{spend}}, pk'_{\text{view}}, \rho', r')$ from the
decrypted opening and accepts only if it equals $\mathsf{cm}_{\text{out}}$. This
binds the encrypted payload to the verified transfer at no additional circuit
cost. Any participant not in $\{B_i, B_j, \mathsf{CB}\}$ holds no wrapped key and
therefore cannot recover $K$, providing 256-bit symmetric security for the
ciphertext $C$.

The payload contains:
$\mathsf{payload} = (\mathsf{sender}, \mathsf{receiver}, v, \mathsf{currency},
\mathsf{timestamp}, \mathsf{reference})$.

\section{ZK Circuit Design}

\subsection{Constraint System}

The ConfidentialTransfer circuit, written in Circom~\cite{circom}, enforces
five classes of constraints:

\begin{enumerate}
  \item \textbf{Spend-key derivation}: $pk_{\text{spend}} =
  \mathsf{DerivePk}(sk_{\text{spend}})$ via scalar multiplication on Baby Jubjub
  followed by a Poseidon compression of the resulting point, so that the
  spending public key committed in the note is bound to the secret key used for
  the nullifier.

  \item \textbf{Input commitment validity}: $\mathsf{cm}_{\text{in}} =
  \mathsf{Poseidon}(v, pk_{\text{spend}}, pk_{\text{view}}, \rho, r)$ is a leaf
  in the Merkle tree rooted at $\mathsf{root}$, verified by a depth-8 MerkleProof
  sub-circuit using Poseidon as the internal hash.

  \item \textbf{Nullifier correctness}: $\mathsf{nf} =
  \mathsf{Poseidon}(\rho, sk_{\text{spend}})$ equals the public input
  $\mathsf{nf}_{\mathsf{pub}}$.

  \item \textbf{Output commitment correctness}: $\mathsf{cm}_{\text{out}} =
  \mathsf{Poseidon}(v', pk'_{\text{spend}}, pk'_{\text{view}}, \rho', r')$ equals
  the public input $\mathsf{cm}_{\mathsf{pub}}$.

  \item \textbf{Value conservation}: $v = v'$.
\end{enumerate}

The circuit has 11,003 wires. The public inputs are the four-element vector
$[\mathsf{root}, \mathsf{nf}_{\mathsf{pub}}, \mathsf{cm}_{\mathsf{pub}},
\mathsf{addr}_{\mathsf{sender}}]$, where the sender address enforces that only
the note owner can initiate the transfer. Because Groth16 verification cost
depends on the public-input count alone, the on-chain verifier is insensitive to
the internal size of the circuit.

\subsection{Poseidon Zeros}

The empty Merkle tree uses Poseidon-iterated zero values at each level.
For depth $d = 8$ with field prime $p$ (BN254 scalar field):
\[
  \mathsf{z}_0 = \mathsf{Poseidon}(0, 0), \quad
  \mathsf{z}_i = \mathsf{Poseidon}(\mathsf{z}_{i-1}, \mathsf{z}_{i-1})
\]
These constants are hardcoded in both the circuit and the on-chain
\texttt{ConfidentialEuroToken} contract to ensure consistency.

\subsection{Incremental Merkle Tree}

The on-chain Merkle tree uses an incremental construction~\cite{tornado2019}
with a \texttt{filledSubtrees[8]} array storing the last left-side node at
each level. Each insertion requires exactly $d = 8$ Poseidon hash calls,
regardless of tree size. A rolling window of the 30 most recent roots is
maintained, allowing proofs generated concurrently to remain valid even when
other notes are inserted between proof generation and proof submission.

\section{Implementation and Evaluation}
\label{sec:evaluation}

\subsection{Deployment Configuration}

The PoC runs on an Oracle Cloud Infrastructure ARM instance (Ampere Altra,
ARM Neoverse N1, 4~OCPUs, 24\,GB RAM, Ubuntu 22.04) — referred to below as
\emph{the PoC hardware} — operating four Hyperledger Besu QBFT validator
nodes as Docker containers. Chain ID~1337, 2-second block period, Cancun
EVM, zero base fee. All external access is via Cloudflare Tunnel.

Three Next.js web applications serve as portals for the central bank, the
central securities depository, and participant banks of the CBDC network. The ZK service
(Node.js, Express) runs as a separate Docker container at a fixed internal
IP and handles all proof generation and cryptographic operations.

\subsection{On-Chain Gas Costs}

Table~\ref{tab:gas} reports gas consumption for each protocol operation,
measured on the PoC network. The higher values compared to
na\"ive estimates reflect three factors: (i) cold SSTORE costs for first-write
slots in the Merkle tree and NoteRegistry arrays; (ii) the ECIES-encrypted
note blob ($\approx$600\,B) written to chain storage; and (iii) the
\texttt{transferFrom} call within \texttt{shield()} that
touches the EuroToken contract storage.

\begin{table}[h]
\centering
\caption{Gas Consumption per Operation (measured on PoC)}
\label{tab:gas}
\small
\begin{tabular}{p{3.1cm}rp{2.1cm}}
\toprule
\textbf{Operation} & \textbf{Gas} & \textbf{Notes} \\
\midrule
\texttt{shield()} & 890,409 & incl.\ approve + storeNote \\
\texttt{confTransfer()} & 1,152,035 & incl.\ verifyProof + storeNote \\
\texttt{unshield()} & 352,249 & incl.\ verifyProof \\
\texttt{EuroToken.\allowbreak approve()} & 46,893 & pre-shield step \\
Groth16 verify & $\approx$220,000 & BN254 pairings (EIP-196/7) \\
\texttt{storeNote()} & $\approx$45,000 & per blob ($\approx$600\,B) \\
\bottomrule
\end{tabular}

{\footnotesize Gas values measured on Hyperledger Besu QBFT, Chain ID 1337,
zero base fee. First-write SSTORE costs dominate cold slots.}
\end{table}

\noindent With zero base fee, gas cost is not a barrier in this permissioned
setting. For public networks, the Groth16 verification cost of $\approx$220k
gas ($\approx$0.0044\,ETH at 20\,Gwei, i.e.\ roughly \$9 at a reference price
of \$2{,}000/ETH) is acceptable for high-value settlement transactions.

\subsection{Proof Generation Performance}

Table~\ref{tab:proof} summarises off-chain proof generation on the PoC hardware,
with software witness generation via \texttt{snarkjs}.

\begin{table}[h]
\centering
\caption{Operation Timing (measured on PoC hardware)}
\label{tab:proof}
\small
\begin{tabular}{p{3.8cm}r}
\toprule
\textbf{Phase} & \textbf{Time} \\
\midrule
\texttt{shield()} end-to-end & $\approx$8\,s \\
\texttt{confTransfer()} end-to-end & $\approx$16\,s \\
\texttt{unshield()} end-to-end & $\approx$16\,s \\
\midrule
Groth16 proof (software) & $\approx$4--12\,s \\
ECIES enc/dec (payload + note) & $< 0.1$\,s \\
NoteRegistry read + decrypt + filter & $\approx$1\,s \\
On-chain verification & $\approx$1\,ms \\
\bottomrule
\end{tabular}

{\footnotesize End-to-end times measured from API call to transaction
confirmed on-chain (block period 2\,s) on the deployment hardware of
\S\ref{sec:evaluation}. The registry row is measured against the owner-keyed
registry of the present implementation (\S\ref{sec:registrylimit}); the
de-indexed design would instead require trial decryption linear in the
network-wide note count. Hardware ZK accelerators reduce proof
generation to $<$1\,s~\cite{cysic2023, ingonyama2023}.}
\end{table}

On the PoC hardware, end-to-end operation times are
8\,s for \texttt{shield}, 16\,s for \texttt{confidentialTransfer}, and 16\,s
for \texttt{unshield}. The dominant cost remains
ZK proof generation; hardware accelerators (e.g., Cysic ZPU, Ingonyama ICICLE
on GPU) reduce Groth16 proof time to under 1\,second for circuits of comparable
size~\cite{cysic2023, ingonyama2023}, making the protocol practical for
production settlement where sub-second latency is required.

\subsection{NoteRegistry Storage Overhead}

Each encrypted note blob is approximately 600 bytes. The gas cost of one
\texttt{storeNote} call is $\approx$45,000 gas, dominated by the cold SSTORE
for the dynamic \texttt{bytes[]} array. For a production-depth tree ($d = 32$)
and a network of 10 banks executing 100 confidential transfers per day, the total on-chain note storage grows by
$\approx$60\,KB/day — negligible for enterprise infrastructure.

\section{Security Analysis}

\subsection{Receiver Confidentiality}

In the protocol as specified, the receiver's address appears only in the
private witness and inside the ECIES-encrypted note blob: it is absent from the
calldata, and the NoteRegistry stores blobs without any owner index
(\S\ref{sec:noteregistry}); \S\ref{sec:registrylimit} records where the current
implementation departs from this. The blob is semantically secure (IND-CCA2) under
the EC-DDH assumption over secp256k1. The commitment $\mathsf{cm}_{\text{out}}$
is computationally binding (Poseidon collision resistance) and hiding (random
$\rho', r'$ blind $pk_{\text{spend},j}$ and $pk_{\text{view},j}$). An adversary learning $\mathsf{cm}_{\text{out}}$
gains no information about $B_j$ beyond what is computable from public data.

\subsection{Amount Confidentiality}

The transferred amount $v$ is a private witness. It is committed to via
Poseidon (one-way, collision resistant) and appears in no public signal.
The circuit constraint $v = v'$ is enforced by the ZK proof; the verifier
accepts only if value is conserved, without learning the value itself.

\subsection{Double-Spend Prevention}

The nullifier $\mathsf{nf} = \mathsf{Poseidon}(\rho, sk_{\text{spend}})$ is
deterministic for a given $(\rho, sk_{\text{spend}})$ pair. The on-chain
\texttt{nullifierSpent} mapping ensures each nullifier is accepted at most once.
Since $\rho$ is unique per note (sampled uniformly at random during note
generation), the nullifier uniquely identifies the note across the system lifetime.

\subsection{Sender Accountability}

The sender address $\mathsf{addr}_i$ is a public input to the circuit and is
verified by the on-chain contract to match \texttt{msg.sender}. A bank
cannot generate a valid proof claiming a different sender without knowledge
of that sender's $sk_{\text{spend}}$. This maintains full on-chain sender
accountability.

\subsection{Soundness}

The Groth16 proof system satisfies knowledge soundness in the generic group
model~\cite{groth2016size}: no computationally bounded prover can produce a
valid proof for a false statement. Within \texttt{confidentialTransfer}, a
prover therefore cannot spend a note absent from the commitment tree, spend a
note it does not own (nullifier correctness requires knowledge of
$sk_{\text{spend}}$), or alter the transferred value, since $v$ is carried
entirely inside commitments and constrained by $v = v'$. These guarantees do
not extend to the transparent boundary operations, where value is bound by
contract-level authorisation rather than by the proof
(\S\ref{sec:valuebinding}).

\section{Discussion and Limitations}

\subsection{Single-Denomination Constraint}

The current circuit enforces $v_{\text{in}} = v_{\text{out}}$: the full note
value must be transferred. A production system would require a two-output
circuit:
\[
  v_{\text{in}} = v_{\text{out},1} + v_{\text{out},2}
\]
where $v_{\text{out},2}$ is a change note returned to the sender. This is a
standard extension in UTXO-based privacy systems~\cite{hopwood2022zcash} and
would approximately double the constraint count.

\subsection{Merkle Tree Capacity}

The PoC fixes the tree depth at $d = 8$, bounding the system to $2^8 = 256$
commitments over its entire lifetime — sufficient for functional validation
but not for production. Raising the depth to $d = 32$ ($4.3 \times 10^9$
notes) costs 32 Poseidon calls per insertion instead of 8; in the circuit,
the Merkle inclusion sub-circuit grows linearly in $d$, so the non-linear
constraint count rises accordingly. Proof size and on-chain verification cost
are unaffected, since the public-input vector is unchanged.

\subsection{Value Binding at the Transparent Boundary}
\label{sec:valuebinding}

The confidential layer conserves value by construction, but the two boundary
operations do not bind the transferred amount to a proof. In \texttt{shield},
the depositing bank supplies the amount and the commitment independently, and
no proof relates the value committed inside $\mathsf{cm}$ to the tokens
actually escrowed. In \texttt{unshield}, the withdrawn amount is a contract
argument and does not appear in the public-input vector, so the proof attests
to the destruction of a note without attesting to its value. A malicious
participant could therefore mint confidential value at the shield boundary or
over-withdraw at the unshield boundary, bounded only by the vault balance.

In the PoC this residual trust is absorbed by the permissioning model: only
authorised institutions can invoke either operation, and both leave a public,
attributable trace. A production deployment must remove the assumption. Two
standard constructions suffice: exposing $v$ as a public input for
\texttt{unshield}, which binds the withdrawn amount to the burnt note at no
additional circuit cost; and, for \texttt{shield}, either a proof of correct
opening of $\mathsf{cm}$ with respect to the deposited amount or homomorphic
value commitments in the style of Zcash~\cite{hopwood2022zcash}, which allow
the balance to be checked without revealing individual values.

\subsection{NoteRegistry Indexing in the Proof of Concept}
\label{sec:registrylimit}

The NoteRegistry design of \S\ref{sec:noteregistry} stores blobs in a single
append-only list carrying no recipient identifier. The current PoC
implementation departs from this design: blobs are held in a mapping keyed by
owner address, the recipient is passed as an explicit argument of
\texttt{confidentialTransfer}, and the storage event carries the owner as an
indexed topic. Receiver confidentiality as defined in \S\ref{sec:model} is
therefore \emph{not} achieved by the current implementation, even though the
amount and the transaction payload remain hidden. Closing the gap requires
dropping the owner argument, replacing owner-keyed reads with recipient-side
trial decryption, and removing the owner from the event signature; the circuit,
the proof system and the on-chain verifier are unaffected.

\subsection{Trusted Setup}

The Groth16 proving key was generated by a single contributor in this PoC.
A production deployment requires a multi-party computation (MPC) ceremony
involving multiple independent parties (analogous to the Zcash Powers of Tau
ceremony~\cite{bowe2017scalable}) to ensure that no single party can produce
false proofs.

\subsection{Viewing Keys}

The current audit model requires the central bank to attempt decryption of
each transaction's wrapped key individually. A Zcash-style viewing key
scheme~\cite{hopwood2022zcash} would derive a single scan key from the
master audit key, enabling efficient batch scanning without per-transaction
effort.

\subsection{Key Derivation and Isolation}

Each participant's spending key, ECIES viewing key, and Ethereum signing key
must be mutually independent and each derived from a wallet-bound secret through
a one-way (hardened) key-derivation function---for example
$sk_{\text{spend}} = \mathsf{KDF}(\sigma) \bmod \ell$, where $\sigma$ is a
domain-separated deterministic signature and $\mathsf{KDF}$ a cryptographic
hash. In particular, the Baby Jubjub spending key must not be produced by
reducing the raw Ethereum private key modulo $\ell$: because the secp256k1 order
$n \approx 2^{256}$ and the subgroup order $\ell \approx 2^{250.6}$ differ only
by $n/\ell \approx 42$, that reduction is invertible in at most $43$ trials
against the public sender address, so a leaked spending key would recover the
Ethereum master key. Hardened derivation confines a key compromise to its own
capability: disclosure of a viewing key then affects confidentiality alone and
does not escalate to spending authority or account control. Our prototype
instantiates $\mathsf{KDF}$ as $\mathsf{SHA\text{-}512}$ applied to a
domain-separation tag concatenated with the wallet-bound secret, reduced modulo
$\ell$. The derivation is off-circuit and leaves the constraint system, the
on-chain verifier, and its gas cost unchanged.

\subsection{Metadata Leakage}

Transaction timing, ciphertext sizes, and block-level co-occurrence of
shield and unshield operations may leak information through traffic analysis.
Mitigation strategies include mandatory delays, dummy transactions, and
ciphertext padding — standard countermeasures studied in the
anonymity literature~\cite{dingledine2004tor}.

\subsection{Single ZK Service}

In the PoC, a single shared ZK service handles proof generation for all banks
and holds all participant key material. As implemented, the sender's service
derives the recipient's spending key when constructing the output note; only
$pk_{\text{spend}}$ is required by the commitment, and the output nullifier is
neither a public input nor needed by the sender, so this is an implementation
shortcut rather than a protocol requirement — but it must be removed before the
per-bank deployment described below. In production, each bank would operate its
own service with isolated key material. The NoteRegistry design is precisely
motivated by this multi-service scenario: note secrets are recoverable from the
chain alone, requiring no cross-service communication.

\section{Conclusion}

We have presented a confidential interbank settlement protocol for permissioned
EVM networks based on the \emph{relaxed sender anonymity} model. The protocol
uses Groth16 zero-knowledge proofs for mathematical settlement verification
without value disclosure, Poseidon-hashed commitments in an incremental Merkle
tree for note management, multi-recipient ECIES encryption for selective
disclosure, and a novel on-chain NoteRegistry for trustless note custody.

The relaxed sender anonymity model is a principled design choice: it preserves
the sender accountability required by financial regulation while providing
strong cryptographic guarantees for receiver identity and transfer amount
confidentiality. The on-chain NoteRegistry eliminates the need for any trusted
intermediary in note distribution, a gap unaddressed by prior work; realising
its receiver-confidentiality property requires the de-indexing described in
\S\ref{sec:registrylimit}, which affects neither the circuit nor the verifier.

The proof-of-concept deployment demonstrates practical feasibility on commodity
hardware. The primary bottleneck — software proof generation, measured at
4--12\,s — is addressable by hardware accelerators, bringing end-to-end
settlement latency below the block time of most enterprise networks.

Future work includes: (i) a two-output circuit enabling partial transfers with
change; (ii) a multi-party trusted setup ceremony; (iii) Zcash-style viewing
keys for efficient audit; and (iv) cross-network interoperability for multi-CBDC
settlement.

\section*{Artifact Availability}

The smart contracts, Circom circuits, deployment scripts and evaluation
tooling supporting this paper are available from the authors on reasonable
request.


\end{document}